\documentclass[aps,prd,reprint,showkeys,nofootinbib,floatfix,superscriptaddress]{revtex4-2}
\usepackage{amsmath,amssymb,bm,graphicx,xcolor,booktabs,array,multirow,siunitx,microtype}
\usepackage[T1]{fontenc}
\usepackage{lmodern}
\usepackage[hidelinks]{hyperref}
\usepackage{enumitem}
\usepackage{capt-of}
\graphicspath{{figures/}}
\newcommand{\dd}{\mathrm{d}}
\newcommand{\cR}{\mathcal{R}}
\newcommand{\cF}{\mathcal{F}}

\definecolor{revisionblue}{RGB}{0,70,160}
\newcommand{\rev}[1]{\textcolor{revisionblue}{#1}}
\colorlet{revisionblue}{black}
\begin{document}

\title{\rev{Kinematic Bounds on Charged Energy Extraction in the Purely Electric Branch of Euler--Heisenberg Type \texorpdfstring{$f(R,T)$}{f(R,T)} Black Holes}}

\author{Anirudh Pradhan}
\email{pradhan.anirudh@gmail.com}
\affiliation{Centre for Cosmology, Astrophysics and Space Science (CCASS), GLA University, Mathura 281406, Uttar Pradesh, India}

\author{K. Ghaderi}
\email{k.ghaderi@iau.ac.ir}
\affiliation{Department of Physics, Mari.C., Islamic Azad University, Marivan, Iran}

\author{M. Zeyauddin}
\email{uddin\_m@rcjy.edu.sa}
\affiliation{Department of General Studies (Mathematics), Jubail Industrial College, Jubail 31961, Saudi Arabia}

\begin{abstract}\color{revisionblue}
We derive and test kinematic upper bounds on charged energy extraction in the purely electric branch of a static $f(R,T)$ black hole family sourced by an Euler--Heisenberg type nonlinear electromagnetic sector. The analysis is restricted to external, minimally coupled probe particles on the fixed background. In this electric solution the field remains Coulombic up to an additive gauge constant; the nonlinear electromagnetic and matter-coupling coefficients therefore affect extraction indirectly through the metric, the event-horizon position, and the global causal structure. We obtain a general local bound retaining nonzero angular momentum and radial velocity of the negative energy fragment, and identify the commonly used zero angular momentum turning-point choice as an upper envelope rather than a generic decay configuration. A co-moving split supplies a locally four momentum-conserving benchmark. We classify horizons with the double root conditions $\mathcal F=\mathcal F'=0$, distinguish asymptotically flat, de Sitter, and anti-de Sitter branches, and impose the corresponding global outward-accessibility criterion. In the de Sitter static patch the electrostatic energy is referenced to the cosmological horizon, so the relevant scale is $Q(1/r_+-1/r_c)$ rather than $Q/r_+$. Reproducible scans compare Reissner--Nordstr\"om, Einstein--Euler--Heisenberg type, $f(R,T)$--Maxwell, and full $f(R,T)$--Euler--Heisenberg type backgrounds. The results show that the branch-referenced horizon potential controls the local upper envelope, whereas the asymptotic structure determines whether an outward trajectory can reach infinity, a cosmological horizon, or only a finite outer turning point. 
\end{abstract}


\maketitle

\section{Introduction}\label{sec:intro}
Energy extraction from black holes provides a sharp way to relate local strong field dynamics to conserved quantities measured in an exterior stationary region. Classical treatments of black hole geometry and test particle motion provide the standard framework used below \cite{Chandrasekhar1983}. In the original Penrose process, rotation creates an ergoregion in which negative Killing energy states are possible, allowing one fragment of a decay to enter the black hole while the other carries more energy than the incident particle \cite{Penrose1969,BardeenPressTeukolsky1972,Wald1974}. Electromagnetic extraction from rotating black holes is represented by the Blandford--Znajek mechanism \cite{BlandfordZnajek1977}. Closely related collisional extensions and their observable efficiency limits have been developed in Refs.~\cite{BSW2009,HaradaKimura2014,Schnittman2014,BertiBritoCardoso2015,HaradaOgasawaraMiyamoto2016,Schnittman2018}, emphasizing the distinction between large local collision energies and energy that can escape to infinity. \rev{A physically distinct plasma mechanism was developed by Comisso and Asenjo: relativistic magnetic reconnection in the ergosphere can produce a negative energy plasma component and an escaping energized component \cite{ComissoAsenjo2021}. These mechanisms require rotation and should not be conflated with the electrostatic process studied below.}

A charged static black hole admits another idealized channel. The canonical energy of a charged particle contains the electrostatic term $q\Phi$, and a fragment whose charge has the opposite sign to the black hole charge can carry negative conserved energy even in a spherically symmetric spacetime. Charged-particle motion in Reissner--Nordstr\"om geometry has been analyzed in detail \cite{PuglieseQuevedoRuffini2011}. \rev{Charged-particle Penrose energetics in Kerr--Newman spacetime were studied by Bhat, Dhurandhar, and Dadhich \cite{BhatDhurandharDadhich1985}, providing a rotating charged benchmark distinct from the static electric process considered here. Tursunov \textit{et al.} studied the electric Penrose process for a weakly charged nonrotating black hole, including ionization and astrophysical charge estimates \cite{Tursunov2021}. Global effects become particularly important in Reissner--Nordstr\"om--(anti-)de Sitter backgrounds, where the asymptotic structure changes both the admissible trajectory domain and the interpretation of energy extraction \cite{Feiteira2024,Feiteira2025}. Charged extraction has also been considered in modified charged geometries \cite{AlloqulovShaymatov2024}.}

\begingroup\color{revisionblue}
Recent complementary studies have examined how deformed black hole geometries affect thermodynamics, shadows, accretion-disk images, greybody factors, scalar perturbations, weak deflection, and related observables \cite{Koam2025,ChaudharyIJTP2025,ChaudharyAscom2026,ChaudharyEPJP2026,ChaudharyPDU2025,ChaudharyJHEAp2025}. These works do not analyze the charged Penrose process itself, but they provide a broader contemporary context for separating local near-horizon modifications from global propagation and observable signatures. The present analysis applies that distinction specifically to charged-particle energetics and outward accessibility.
\endgroup

The present work asks a narrower question: how do the metric deformations of the purely electric black hole solution in $f(R,T)=R+2\beta T$ gravity coupled to an Euler--Heisenberg type nonlinear electromagnetic Lagrangian modify rigorous test particle bounds? The background was derived by Liang, Tao, and Yang \cite{LiangTaoYang2026}, building on $f(R,T)$ gravity \cite{Harko2011} and nonlinear electrodynamics \cite{HeisenbergEuler1936,Schwinger1951,Dunne2012,HardingLai2006,MarklundShukla2006,YajimaTamaki2001,Ruffini2013}. Two qualifications are essential.

\begingroup\color{revisionblue}
First, in the purely electric branch used here the electromagnetic invariant $G$ vanishes, so the coefficient of $G^2$ does not enter. The electric field is Coulombic and the gauge potential differs between exterior branches only by an additive constant. Consequently, the nonlinear coefficient affects the charged-particle bound through the geometry and the horizon location, not through a new radial dependence of the Lorentz-force potential. We therefore use the term \emph{Euler--Heisenberg type}. In QED the quadratic coefficients are fixed combinations of the fine-structure constant and the electron mass and the truncated effective action is restricted to the weak-field regime. By contrast, the coefficients used below are the mass-normalized numerical parameters of the adopted solution family. The broad scan is therefore a phenomenological solution-space survey rather than a scan of laboratory QED coefficients.

Second, the calculation is a probe analysis. In generic $f(R,T)$ models the matter stress tensor need not be covariantly conserved and an effective extra force can arise for matter included in the modified field equations \cite{Harko2011,LiangTaoYang2026}. Here the fragments are external, minimally coupled test particles whose stress tensor is neglected in the background equations. Their motion is therefore governed by the standard Lorentz force on a fixed geometry. This assumption must be retained when interpreting all results.
\endgroup

\rev{The analysis retains the angular momentum and instantaneous radial velocity of the infalling fragment and proves that the zero angular momentum turning-point choice defines an upper envelope. It also includes a fully four momentum-conserving co-moving split as a locally admissible benchmark, while global outward accessibility is tested separately. Event and cosmological horizons are classified explicitly and the extremal boundaries are determined from simultaneous double-root conditions. Distinct global accessibility criteria are formulated for flat, de Sitter (dS), and anti-de Sitter (AdS) branches. Explicit numerical algorithms, convergence tests, data tables, and direct benchmark comparisons are provided to isolate the effects of the nonlinear and matter-coupling coefficients.}

The paper is organized as follows. Section~\ref{sec:background} defines the mass normalized electric solution and its causal structure. Section~\ref{sec:dynamics} develops charged-particle motion and the local extraction bounds. Section~\ref{sec:global} states the global accessibility conditions. Section~\ref{sec:numerics} documents the reproducible numerical procedure. Section~\ref{sec:results} presents the benchmark and parameter-space results. Section~\ref{sec:validity} discusses physical validity and limitations, and Sec.~\ref{sec:conclusions} summarizes the conclusions.

\section{Electric background and causal structure}\label{sec:background}
We consider
\begin{equation}
 \dd s^2=-\cF(r)\dd t^2+\frac{\dd r^2}{\cF(r)}+r^2(\dd\theta^2+\sin^2\theta\,\dd\phi^2),
 \label{eq:metric}
\end{equation}
with the purely electric solution reported in Ref.~\cite{LiangTaoYang2026}. \begingroup\color{revisionblue}
Its electromagnetic sector is based on the quadratic model $\mathcal L_{\rm NLED}=(4\pi)^{-1}(-F+a_{0}F^2+b_{0}G^2)$, where $F=F_{\mu\nu}F^{\mu\nu}/4$ and $G=F_{\mu\nu}{}^{\star}F^{\mu\nu}/4$. To make the numerical conventions explicit, we set $x=r/M$ and $q=Q/M$, convert the coefficients appearing in the solution to the same $M=1$ normalization, denote the resulting numerical coefficients by $\bar a$ and $\bar\beta$, and suppress the bars below. Thus, values such as $a=20$ or $60$ label this normalized solution family; they are not universal QED Wilson coefficients and cannot be transferred to another mass scale or compared with laboratory bounds without restoring the original dimensional normalization. If the quadratic Lagrangian is interpreted as a truncated QED effective action, one must additionally impose the weak-field conditions $|a_0F|\ll1$ and $|b_0G|\ll1$. The broad maps presented here are not claimed to remain inside that perturbative QED domain at every point; they are phenomenological maps of the adopted quadratic nonlinear-electrodynamic geometry. \endgroup
\begingroup\color{revisionblue}
The metric function is
\begin{equation}
 \cF(x)=1-\frac{2}{x}+C_2\frac{q^2}{x^2}+C_6\frac{q^4}{x^6}+A_8x^2,
 \label{eq:F}
\end{equation}
where
\begin{align}
 A_8&=\frac{\beta}{3\pi},\nonumber\\
 C_2&=1+\beta\left(\frac{a}{\pi}+\frac{3}{2\pi}\right),\nonumber\\
 C_6&=a\left(\frac{1}{30}+\frac{\beta}{60\pi}\right).
 \label{eq:coeff}
\end{align}
\endgroup
\begingroup\color{revisionblue}
The electric field fixes the radial dependence of the potential, while an additive constant must be chosen consistently with the global exterior. We write
\begin{align}
 A_t^{\rm(ref)}(x)&=-q\left(\frac{1}{x}-\frac{1}{x_{\rm ref}}\right),\nonumber\\
 \Phi_{\rm ref}(x)&\equiv-A_t^{\rm(ref)}(x)
 =q\left(\frac{1}{x}-\frac{1}{x_{\rm ref}}\right),
 \label{eq:potential}
\end{align}
where $x_{\rm ref}=\infty$ for the asymptotically flat and AdS branches. In the dS static patch, spatial infinity is not part of the exterior domain, and we set $x_{\rm ref}=x_c$, so that $\Phi_{\rm ref}(x_c)=0$. This convention makes the dS energy transfer depend on the potential difference between the black hole and cosmological horizons. Because $G=0$ in this branch, the coefficient multiplying $G^2$ in the nonlinear electromagnetic Lagrangian is absent from Eqs.~\eqref{eq:F}--\eqref{eq:potential}.
\endgroup

The $A_8x^2$ term controls the large radius behavior. The branch $\beta=0$ is asymptotically flat, $\beta<0$ has dS asymptotics, and $\beta>0$ has AdS asymptotics. Multiplication of $\cF=0$ by $x^6$ gives
\begin{equation}
 P(x)=A_8x^8+x^6-2x^5+C_2q^2x^4+C_6q^4=0.
 \label{eq:horizonpoly}
\end{equation}
\begingroup\color{revisionblue}
For $\beta<0$, a nonextremal charged dS black hole has roots
\begin{equation}
 x_-<x_+<x_c,
 \label{eq:dSroots}
\end{equation}
where $x_+$ is the black hole event horizon and $x_c$ is the cosmological horizon. The largest positive root is therefore \emph{not} the event horizon. For $\beta\ge0$, a nonextremal charged black hole has an inner root and an outer event root $x_+$. Degeneracy is not inferred from the number of numerically distinct roots; it is determined by
\begin{equation}
 P(x_e)=0,\qquad P'(x_e)=0.
 \label{eq:double}
\end{equation}
Writing $y=q^2$, the derivative condition gives
\begin{equation}
 y=\frac{10x_e-6x_e^2-8A_8x_e^4}{4C_2},
 \label{eq:extremaly}
\end{equation}
provided $C_2\ne0$, and substitution into Eq.~\eqref{eq:horizonpoly} yields a one dimensional semianalytic extremality equation. These relations independently validate the root-classification boundary used in the numerical maps.
\endgroup

\section{Charged-particle dynamics and local bounds}\label{sec:dynamics}
An external probe particle of mass $m$ and charge $q_p$ is described by the standard minimally coupled action in stationary curved spacetime,
\begin{equation}
 S_p=-m\int \dd s+q_p\int A_\mu\dd x^\mu.
 \label{eq:actionparticle}
\end{equation}
For equatorial motion, the Killing energy and angular momentum are
\begin{equation}
 E=m\cF\dot t+q_p\Phi,\qquad L=mr^2\dot\phi.
 \label{eq:EL}
\end{equation}
The future-directed condition outside an event horizon is
\begin{equation}
 E-q_p\Phi>0.
 \label{eq:forward}
\end{equation}
Defining $\varepsilon=E/m$, $e=q_p/m$, and $j=L/(mM)$, the radial equation becomes
\begin{equation}
 \dot x^2=\cR(x)\equiv
 \left[\varepsilon-e\Phi_{\rm ref}(x)\right]^2-
 \cF(x)\left(1+\frac{j^2}{x^2}\right).
 \label{eq:radial}
\end{equation}

Consider an instantaneous split $0\rightarrow1+2$ at $x=x_s>x_+$. Conservation of canonical energy, electric charge, and azimuthal angular momentum gives
\begin{equation}
 E_0=E_1+E_2,\quad q_0=q_1+q_2,\quad L_0=L_1+L_2.
 \label{eq:conserved}
\end{equation}
\rev{A physical local decay must additionally satisfy four momentum conservation and the mass-shell constraints. The most general statements below are therefore upper bounds conditioned on the existence of a compatible local split.}

\begingroup\color{revisionblue}
For fragment 2, the positive root of Eq.~\eqref{eq:radial} at the split gives
\begin{equation}
 E_2=q_2\Phi_s+m_2
 \sqrt{(u_2^r)^2+\cF_s\left(1+\frac{j_2^2}{x_s^2}\right)},
 \label{eq:E2general}
\end{equation}
where $u_2^r=\dot x_2$, $j_2=L_2/(m_2M)$, $\Phi_s=\Phi_{\rm ref}(x_s)$, and $\cF_s=\cF(x_s)$. For $q>0$, negative energy requires $q_2<0$. We define
\begin{equation}
 \chi_2=\frac{|q_2|}{E_0},\qquad
 \mu_2=\frac{m_2}{E_0},\qquad
 w_2=|u_2^r|,
 \label{eq:ratios}
\end{equation}
and distinguish the escaping amplification factor $\mathcal A=E_1/E_0$ from the net extraction efficiency
\begin{equation}
 \eta_{\rm ext}\equiv\frac{E_1-E_0}{E_0}=-\frac{E_2}{E_0}=\mathcal A-1.
 \label{eq:effdef}
\end{equation}
Equations~\eqref{eq:E2general} and \eqref{eq:effdef} imply
\begin{equation}
 \eta_{\rm ext}^{\rm ub}(x_s)=
 \chi_2\Phi_{\rm ref}(x_s)-\mu_2
 \sqrt{w_2^2+\cF_s\left(1+\frac{j_2^2}{x_s^2}\right)}.
 \label{eq:generalbound}
\end{equation}
The corresponding local negative energy condition is
\begin{equation}
 \chi_2\Phi_{\rm ref}(x_s)>
 \mu_2\sqrt{w_2^2+\cF_s\left(1+\frac{j_2^2}{x_s^2}\right)}.
 \label{eq:negative}
\end{equation}
At fixed $x_s$, every nonzero $w_2$ or $|j_2|$ lowers the bound. Thus, $w_2=j_2=0$ is mathematically the upper envelope of this family, not a statement that all physical decays occur at a radial turning point with zero angular momentum. The horizon limit is
\begin{align}
 \eta_{\rm H}^{\rm ub}
 &=\lim_{x_s\to x_+^+}\eta_{\rm ext}^{\rm ub}
 =\chi_2\Delta\Phi_+-\mu_2w_2,\nonumber\\
 \Delta\Phi_+&\equiv\Phi_{\rm ref}(x_+).
 \label{eq:horizonbound}
\end{align}
For $w_2=0$, this is a supremum reached only in the limiting sequence $x_s\to x_+^+$ unless a separate physical cutoff is supplied.
\endgroup

For the turning-point envelope $w_2=j_2=0$, Eq.~\eqref{eq:negative} can be squared without changing its sign content and the outer boundary of the negative energy region is an exact root of
\begin{align}
 0={}&A_8x^8+\left(1-\kappa^2q^2x_{\rm ref}^{-2}\right)x^6\nonumber\\
 &+\left(-2+2\kappa^2q^2x_{\rm ref}^{-1}\right)x^5
 +\left(C_2-\kappa^2\right)q^2x^4+C_6q^4,\nonumber\\
 &\hspace{2.5em}\kappa\equiv\frac{\chi_2}{\mu_2}.
 \label{eq:negpoly}
\end{align}
We define $\Delta x_{\rm neg}=x_{\rm neg}-x_+$ using the first such root outside $x_+$, thereby selecting the negative energy layer connected to the black hole horizon. A disconnected layer near a dS cosmological horizon is not counted as an event-horizon extraction region.

\begingroup\color{revisionblue}
\subsection{Four momentum-conserving benchmark}\label{subsec:comoving}
To show that the upper envelope analysis is not the only available construction, consider a neutral parent, $q_0=0$, that splits into $q_1=-q_2$ with both fragments initially co-moving with the parent. Let $m_1=(1-\lambda)m_0$ and $m_2=\lambda m_0$. Then
\begin{equation}
 p_0^\mu=m_0u^\mu=p_1^\mu+p_2^\mu,
 \label{eq:fourmomentum}
\end{equation}
so four momentum and charge are conserved exactly at the event. Since $E_0=m_0\cF u^t$ and $E_2=\lambda E_0-|q_2|\Phi_s$, the net efficiency is
\begin{equation}
 \eta_{\rm cm}(x_s)=\chi_2\Phi_{\rm ref}(x_s)-\lambda,
 \qquad
 \eta_{\rm cm,H}=\chi_2\Delta\Phi_+-\lambda.
 \label{eq:comoving}
\end{equation}
This benchmark is more restrictive than Eq.~\eqref{eq:horizonbound} for $w_2=0$ and is manifestly compatible with local four momentum conservation. It establishes local decay admissibility only; any chosen outgoing fragment must still satisfy the global conditions of Sec.~\ref{sec:global}. It is used below with $\lambda=0.1$.
\endgroup

\section{Global outward accessibility}\label{sec:global}
\begingroup\color{revisionblue}
A local negative energy split is insufficient: fragment 1 must remain future directed and radially allowed throughout the relevant exterior domain. With
\begin{equation}
 \varepsilon_1=\frac{E_1}{m_1},\qquad
 e_1=\frac{q_1}{m_1},\qquad
 j_1=\frac{L_1}{m_1M},
\end{equation}
we require
\begin{align}
 \cR_1(x)={}&\bigl[\varepsilon_1-e_1\Phi_{\rm ref}(x)\bigr]^2\nonumber\\
 &-\cF(x)\left(1+\frac{j_1^2}{x^2}\right)\ge0,
 \label{eq:R1}\\
 \varepsilon_1-e_1\Phi_{\rm ref}(x)&>0.
 \label{eq:forward1}
\end{align}
The physical endpoint depends on $\beta$.

For $\beta=0$, escape to spatial infinity requires Eqs.~\eqref{eq:R1}--\eqref{eq:forward1} for all $x\ge x_s$ and $\varepsilon_1\ge1$. For $\beta<0$, the static exterior ends at $x_c$; outward accessibility means propagation from $x_s$ to the cosmological horizon with no intervening turning point. It is not described as escape to spatial infinity. For $\beta>0$, $\cF\sim A_8x^2$, and a massive particle with finite Killing energy has $\cR_1\to-\infty$. It therefore possesses a finite outer turning point and cannot reach the AdS boundary. In that branch we report only finite radius outward propagation and do not label the local horizon bound an asymptotic extraction efficiency.

For reproducibility, turning points are found from an exact polynomial rather than from a finite-window visual test. Defining the gauge-shifted constant
\begin{equation}
 \bar\varepsilon_1\equiv\varepsilon_1+e_1q/x_{\rm ref},
 \label{eq:epsbar}
\end{equation}
Eq.~\eqref{eq:R1} has the same polynomial form as the Coulomb-gauge expression with $\varepsilon_1$ replaced by $\bar\varepsilon_1$. Multiplication by $x^8$ gives
\begin{align}
 x^8\cR_1={}&-A_8x^{10}+
 (\bar\varepsilon_1^2-1-j_1^2A_8)x^8
 +(2-2\bar\varepsilon_1e_1q)x^7\nonumber\\
 &+[(e_1^2-C_2)q^2-j_1^2]x^6+2j_1^2x^5
 -j_1^2C_2q^2x^4\nonumber\\
 &-C_6q^4x^2-j_1^2C_6q^4.
 \label{eq:radialpoly}
\end{align}
All positive real roots beyond $x_s$ are computed and filtered by the appropriate outer domain. The single zero of Eq.~\eqref{eq:forward1} is checked independently.
\endgroup

\section{Numerical implementation}\label{sec:numerics}
\rev{All numerical results follow directly from Eqs.~\eqref{eq:F}--\eqref{eq:radialpoly}.}

\begingroup\color{revisionblue}
\begin{table*}[t]
\caption{\rev{Numerical protocol. The default particle ratios normalize the kinematics while the overall probe scale remains arbitrary and can be chosen to satisfy Eq.~\eqref{eq:probeconditions}.}}
\label{tab:protocol}
\begin{ruledtabular}
\begin{tabular}{ll}
Quantity & Value or method \\
\hline
Parameter grid & $q\in[0.10,1.15]$ (141 points), $a\in[0,60]$ (121 points) \\
Matter coupling & $\beta=-0.02,0,+0.02$ \\
Default fragment 2 & $\chi_2=1$, $\mu_2=0.10$, $j_2=0$, $w_2=0$ \\
Escape benchmark & $\mu_1=0.50$, $j_1=0$ unless varied \\
Horizon roots & all positive real roots of Eq.~\eqref{eq:horizonpoly} \\
Root acceptance & relative imaginary part $<10^{-8}$ \\
Root merging & relative separation $<2\times10^{-6}$ \\
Simple-root refinement & bracketed Brent solve, tolerance $10^{-13}$ \\
Negative energy boundary & exact polynomial Eq.~\eqref{eq:negpoly} \\
Escape turning points & exact polynomial Eq.~\eqref{eq:radialpoly} \\
Extremal boundary & root-class bisection, checked by Eqs.~\eqref{eq:double}--\eqref{eq:extremaly}
\end{tabular}
\end{ruledtabular}
\end{table*}
\endgroup

The exported extremal-boundary audit contains the numerical continuation point, the associated degenerate radius, and the residuals of the two independent conditions. Across the full three-branch audit, the largest residuals are $|\cF|<6.1\times10^{-14}$ and $|\cF'|<3.5\times10^{-6}$; the remaining derivative residual reflects the finite bisection location of the root-class transition rather than a fitted boundary.

\begingroup\color{revisionblue}
The horizon limit convergence was tested at five representative parameter points. For $x_s=x_+(1+\epsilon)$, the largest absolute difference between Eq.~\eqref{eq:generalbound} and its analytic horizon limit over those points decreases as shown in Table~\ref{tab:conv}. The noninteger rate is expected because the leading correction near a simple horizon contains $\sqrt{\cF}\propto\sqrt{x_s-x_+}$.

\begin{table}[t]
\caption{\rev{Maximum absolute horizon-regularization error over five test points: $(q,a,\beta)=(0.70,20,-0.02)$, $(0.70,20,0)$, $(0.70,20,0.02)$, $(0.85,5,0)$, and $(0.50,60,0.02)$.}}
\label{tab:conv}
\centering
\small
\setlength{\tabcolsep}{4pt}
\begin{tabular}{@{}c c@{}}
\toprule
$\epsilon$ & $\max|\eta_{\rm ext}^{\rm ub}[x_+(1+\epsilon)]-\eta_{\rm H}^{\rm ub}|$\\
\midrule
$10^{-2}$ & $1.36\times10^{-2}$\\
$10^{-3}$ & $3.27\times10^{-3}$\\
$10^{-4}$ & $9.72\times10^{-4}$\\
$10^{-5}$ & $3.01\times10^{-4}$\\
$10^{-6}$ & $9.47\times10^{-5}$\\
\bottomrule
\end{tabular}
\end{table}
\endgroup

\section{Results}\label{sec:results}
\subsection{Metric profiles and causal domain}
\rev{Figure~\ref{fig:metricprofiles} shows how $a$, $q$, and $\beta$ deform $\mathcal F(x)$, while open markers identify the event horizons used in the extraction analysis. The curves are displayed across the root region only to visualize the causal transitions; the physical black hole exterior begins at $x_+$. The panel confirms that the higher inverse-power correction is concentrated near the horizon and that the matter coupling changes both the near-horizon profile and the large-radius branch.}

\begingroup\color{revisionblue}
\begin{figure*}[t]
\includegraphics[width=0.9\textwidth]{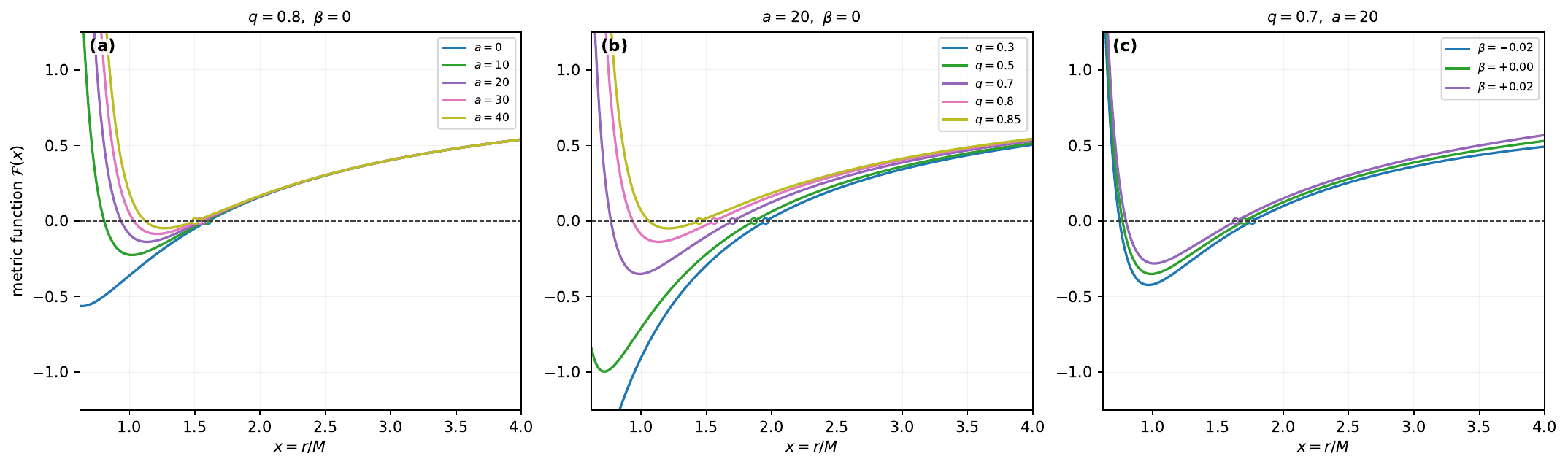}
\caption{\rev{Metric function of the purely electric branch. (a) Variation with the mass-normalized nonlinear coefficient $a$ at fixed $q=0.8$ and $\beta=0$; (b) variation with $q$ at fixed $a=20$ and $\beta=0$; (c) variation with $\beta$ at fixed $q=0.7$ and $a=20$. Open circles mark black hole event horizons; an open square would mark a cosmological horizon if it entered the displayed interval. The dashed line is $\mathcal F=0$. The plotted interior portions are included only to expose the root structure.}}
\label{fig:metricprofiles}
\end{figure*}
\endgroup

Figure~\ref{fig:causal} provides the quantitative causal counterpart. Panel (a) shows the limiting charge obtained from the simultaneous double root conditions \eqref{eq:double}--\eqref{eq:extremaly}; black hole configurations lie below the corresponding curve. Panel (b) shows the event horizon at $a=20$. The dS cosmological horizon is plotted on its own right axis because it is an order of magnitude larger than $x_+$. This representation avoids identifying a single numerical root with a degenerate horizon and displays the causal boundary quantitatively.

\begingroup\color{revisionblue}
\begin{figure*}[t]
\includegraphics[width=0.8\textwidth]{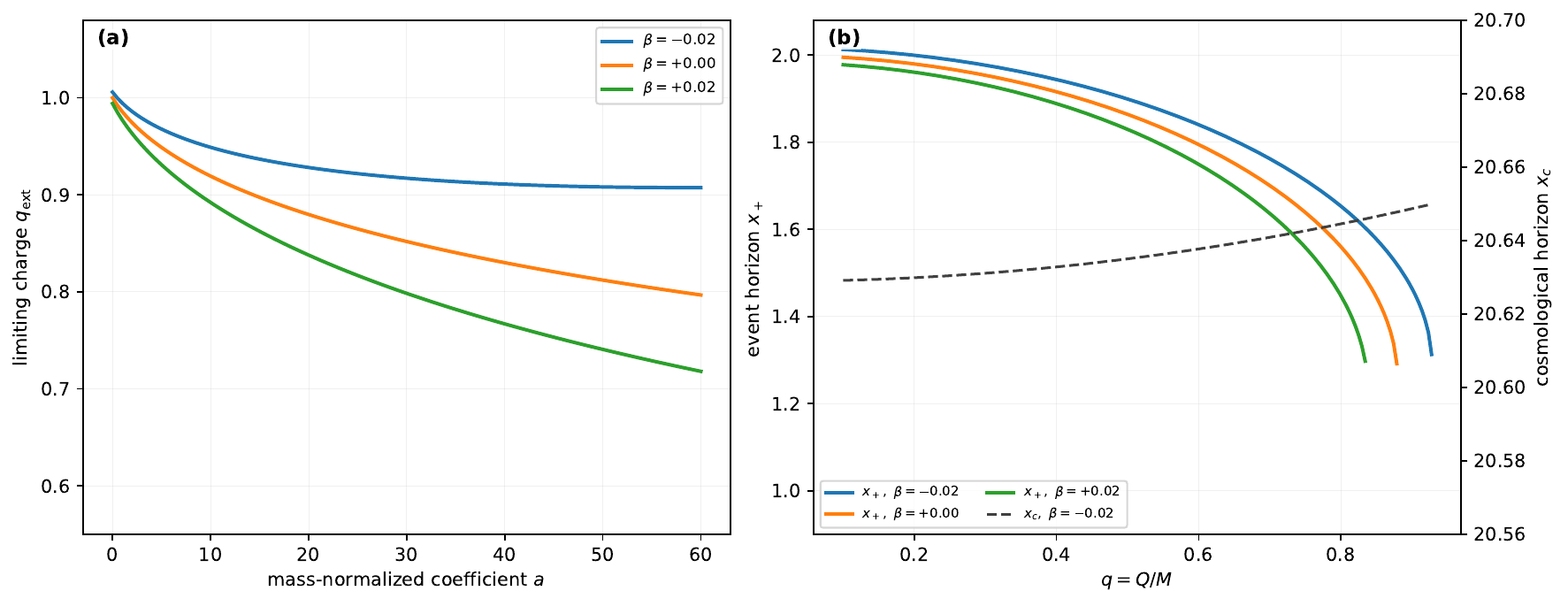}
\caption{\rev{Causal and extremal diagnostics. (a) Limiting charge $q_{\rm ext}(a)$ determined by $\mathcal F=\mathcal F'=0$ for the three representative matter couplings; black hole configurations occupy the lower-charge side. (b) Event-horizon radius at $a=20$; the dashed curve and right axis show the dS cosmological horizon. Curves terminate at the corresponding extremal boundary.}}
\label{fig:causal}
\end{figure*}
\endgroup

\subsection{Direct benchmark decomposition}
\begingroup\color{revisionblue}
Figure~\ref{fig:benchmarks} compares four nested backgrounds: RN ($a=\beta=0$), Einstein--Euler--Heisenberg type ($a=20$, $\beta=0$), $f(R,T)$--Maxwell ($a=0$, $\beta=0.02$), and the full model ($a=20$, $\beta=0.02$). The dS case is not mixed into this benchmark plot because its energy is referenced to the cosmological horizon rather than to infinity; it is included consistently in the branch-dependent maps below.

The event-horizon shift in panel (a) maps directly into the branch-referenced horizon potential in panel (b). Panel (c) shows the locally four momentum-conserving co-moving benchmark \eqref{eq:comoving} with $\chi_2=1$ and $\lambda=0.1$. Panel (d) gives the absolute potential shift relative to RN, avoiding the artificial divergence of a percentage comparison near a branch endpoint. The model dependence is predominantly a geometric renormalization of the RN-like potential scale through $x_+$.
\endgroup

\begingroup\color{revisionblue}
\begin{figure*}[t]
\includegraphics[width=0.8\textwidth]{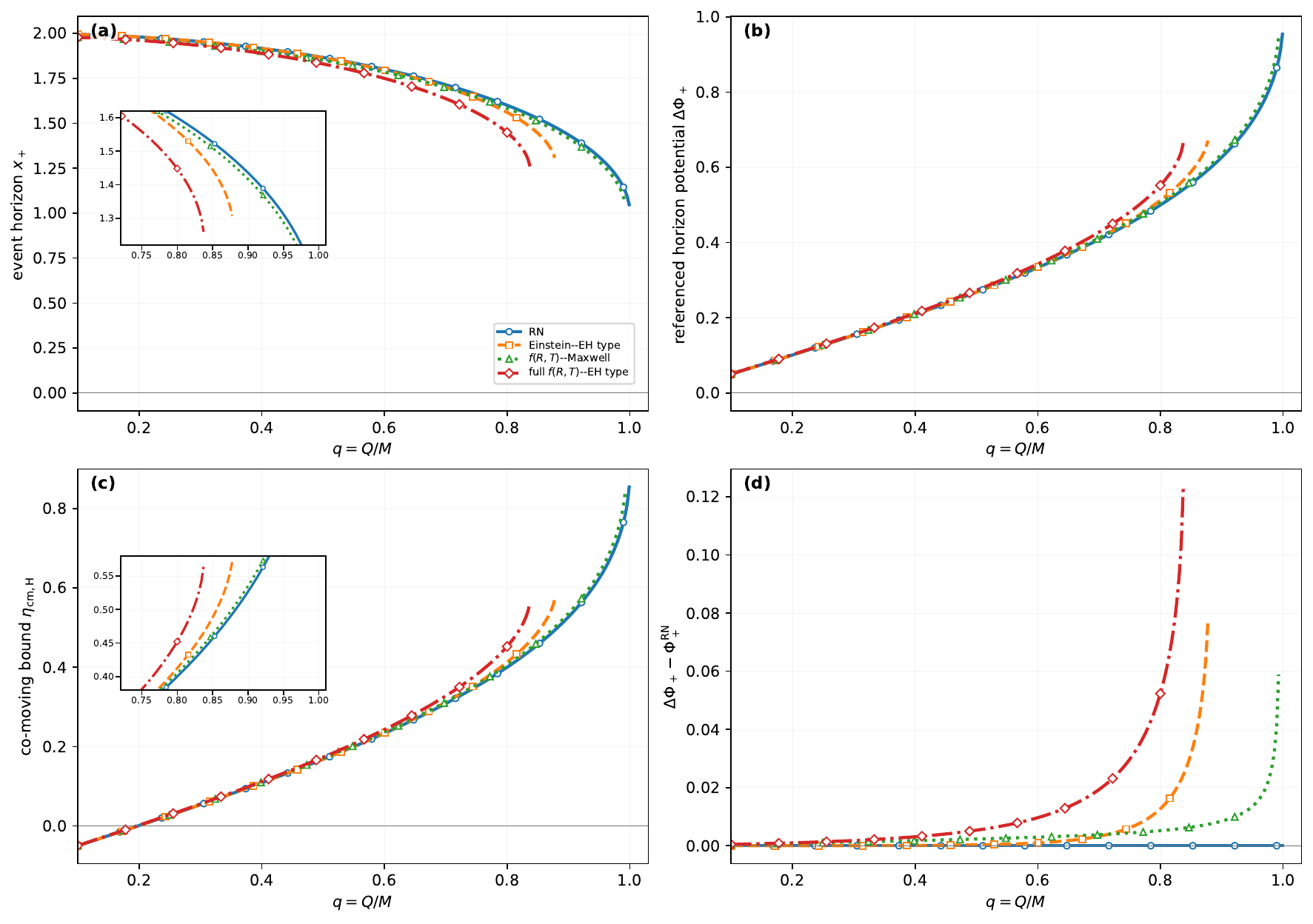}
\caption{\rev{Direct benchmark decomposition for RN ($a=\beta=0$), Einstein--Euler--Heisenberg Type ($a=20$, $\beta=0$), $f(R,T)$--Maxwell ($a=0$, $\beta=0.02$), and the full model ($a=20$, $\beta=0.02$). Panels show (a) the event horizon, (b) the referenced horizon potential $\Delta\Phi_+$, (c) the locally four momentum-conserving co-moving bound with $\lambda=0.1$, and (d) the absolute potential shift relative to RN. Insets in panels (a) and (c) resolve the near-endpoint separation. Curves terminate when the corresponding black hole branch ends.}}
\label{fig:benchmarks}
\end{figure*}
\endgroup

\subsection{Negative energy region and local horizon envelope}
\begingroup\color{revisionblue}
Figure~\ref{fig:maps} combines the two principal parameter diagnostics. The upper row shows the connected negative energy width for the default particle ratios. This width depends on $\chi_2/\mu_2$ and is not a background invariant. The lower row shows the local horizon envelope. In the dS panels both quantities use the cosmological-horizon reference of Eq.~\eqref{eq:potential}; in the flat and AdS panels the reference is infinity. White parameter regions have no black hole event horizon. 
\endgroup

\begingroup\color{revisionblue}
\begin{figure*}[t]
\includegraphics[width=0.8\textwidth]{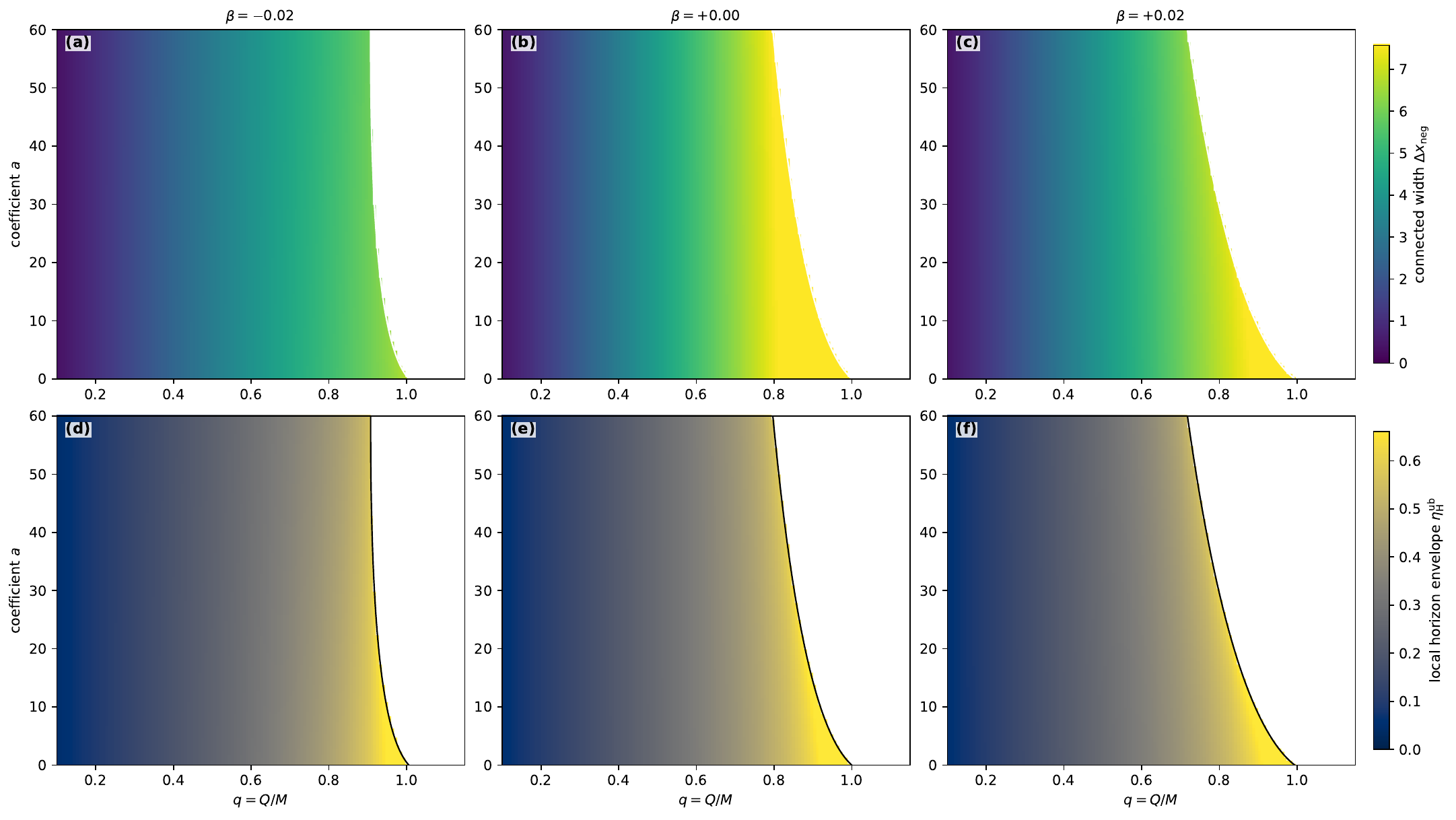}
\caption{\rev{Branch-consistent parameter maps for $\chi_2=1$, $\mu_2=0.1$, and $j_2=w_2=0$. Panels (a)--(c): width $\Delta x_{\rm neg}$ of the negative energy layer connected to the event horizon. Panels (d)--(f): local horizon envelope $\eta_{\rm H}^{\rm ub}=\chi_2\Delta\Phi_+$. Columns correspond to $\beta=-0.02,0,+0.02$. The overlaid curves are the extremal boundaries. The dS column uses $\Delta\Phi_+=q(1/x_+-1/x_c)$.}}
\label{fig:maps}
\end{figure*}
\endgroup

\subsection{Sensitivity and global accessibility}
\begingroup\color{revisionblue}
Panels (a) and (b) of Fig.~\ref{fig:sensitivity} quantify the idealization behind the local envelope. Nonzero $j_2$ lowers the bound away from the horizon, and nonzero radial speed lowers both the curve and its limiting value by $\mu_2w_2$. This verifies that the turning-point, zero angular momentum configuration is a controlled upper limit rather than a representative generic split.
\endgroup

Panel (c) demonstrates the branch dependence of outward accessibility at $(q,a)=(0.7,20)$ and $x_s=x_+(1+10^{-4})$. The radial function remains positive from the split to infinity in the flat branch and to the cosmological horizon in the dS branch, whereas the AdS trajectory crosses zero at a finite radius. Panel (d) shows the continuous angular-momentum dependence of the first AdS turning point. At $j_1=0,2,5$, the roots are $57.59$, $57.56$, and $57.35$, respectively. These are roots of Eq.~\eqref{eq:radialpoly}, not endpoints of the plotting window.

\begingroup\color{revisionblue}
\begin{figure*}[t]
\centering
\includegraphics[width=0.8\textwidth]{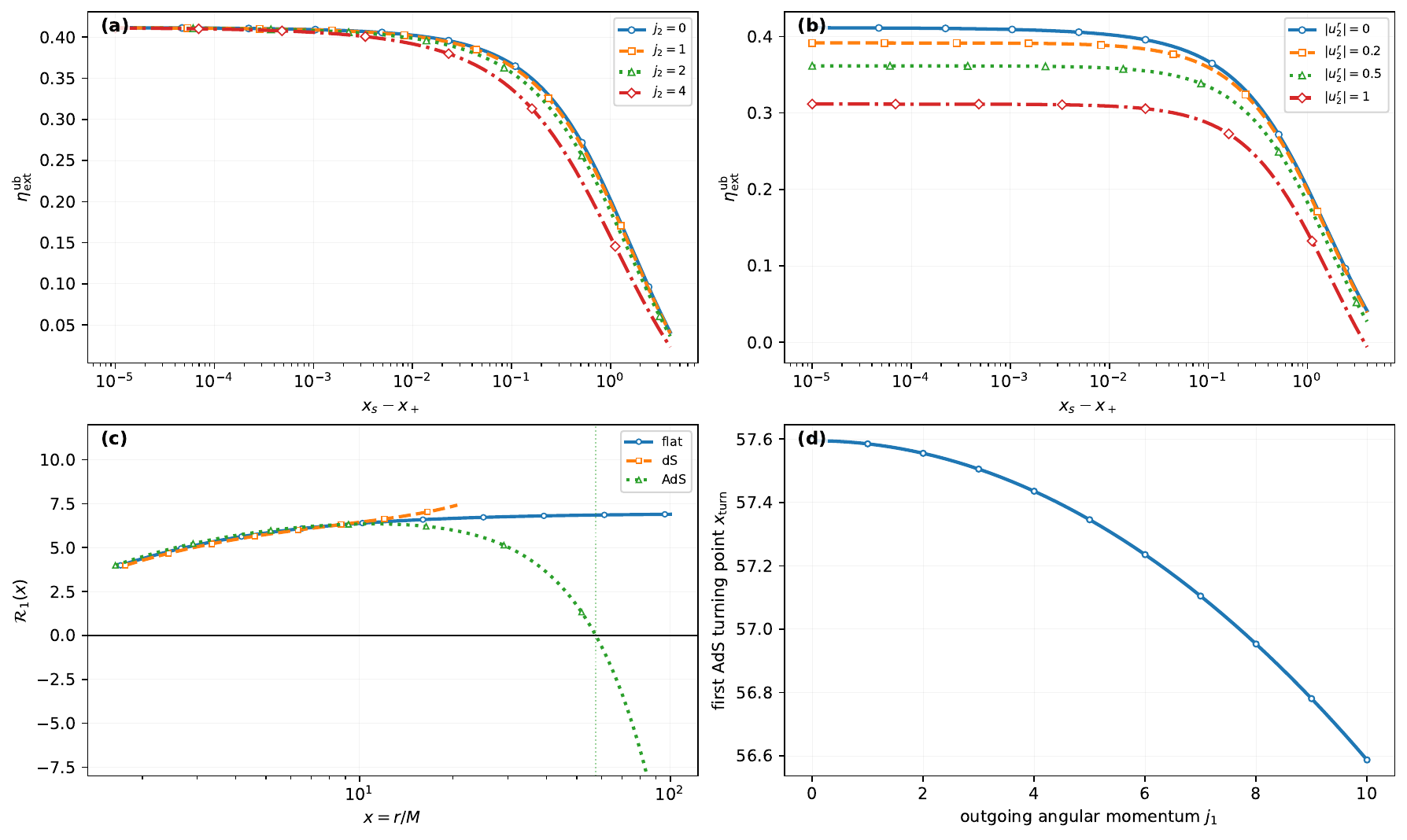}
\caption{\rev{Sensitivity and global accessibility at $q=0.7$, $a=20$. (a) Dependence of the local bound on fragment-2 angular momentum for $\beta=0$ and $w_2=0$. (b) Dependence on radial speed for $j_2=0$. (c) Fragment-1 radial function for the flat, dS, and AdS branches; the dS curve uses the cosmological-horizon potential reference. (d) First AdS turning point as a function of outgoing angular momentum. Default ratios are given in Table~\ref{tab:protocol}.}}
\label{fig:sensitivity}
\end{figure*}
\endgroup

\section{Domain of validity and physical interpretation}\label{sec:validity}
\subsection{Probe and backreaction conditions}
\begingroup\color{revisionblue}
The parameter maps must first be read with the model convention stated in Sec.~\ref{sec:background}. They survey the normalized quadratic nonlinear-electrodynamic solution and are not direct QED coefficient constraints. A strict QED effective-action interpretation would require a separate weak-field mask based on $|a_0F|\ll1$ (and $|b_0G|\ll1$, although $G=0$ here); points outside that domain retain meaning only as configurations of the phenomenological quadratic model.

The dimensionless ratios used in the maps do not fix the absolute scale of the split. The test particle approximation requires an overall normalization such that
\begin{equation}
 \frac{E_i}{M}\ll1,\qquad \frac{m_i}{M}\ll1,\qquad
 \frac{|q_i|}{|Q|}\ll1,
 \label{eq:probeconditions}
\end{equation}
while the ratios $m_i/E_0$ and $|q_i|/E_0$ entering the kinematic bound are held fixed. These conditions ensure that a single event does not appreciably change the mass, charge, or metric of the background. Repeated extraction would require an evolution model for $M$ and $Q$ and cannot be inferred by iterating the fixed background bound without charge depletion.

Self-force and radiation reaction are omitted. Their relative importance increases for large specific charge, strong acceleration, and trajectories forced to remain extremely close to the horizon. Pair creation can screen a sufficiently strong electric field, and accelerated charged particles can modify the local electromagnetic environment. Plasma currents, accretion, and selective capture can either generate a small effective charge or neutralize it. These effects set physical cutoffs on the formal limit $x_s\to x_+$. The present result is therefore conditional: it states the maximum Killing-energy transfer allowed by the fixed background probe equations, not the efficiency of a self consistent plasma discharge.
\endgroup

\subsection{Astrophysical charge and omitted losses}
\begingroup\color{revisionblue}
Astrophysical black holes are expected to be close to neutral because ambient plasma preferentially accretes the opposite sign of charge. Weak effective charges may nevertheless arise from selective accretion or magnetospheric equilibrium, and this motivates studying the electric Penrose process as a diagnostic toy model \cite{Wald1974Magnetic,Tursunov2021}. Tursunov \textit{et al.} quote characteristic charges of approximately $10^{11}$--$10^{18}(M/M_\odot)$ statcoulomb, whereas their gravitationally extremal scale is of order $10^{30}(M/M_\odot)$ statcoulomb \cite{Tursunov2021}. In the corresponding geometric normalization this is only an order of magnitude estimate, $q\sim10^{-19}$--$10^{-12}$. The formal scans up to $q\sim1$ therefore explore the geometry near its charged extremal boundary; they are not asserted to be observationally typical charge to mass ratios.

\rev{No physical energy table is inferred from $\Delta\Phi_+$. In geometrized RN normalization, converting $Q/r_+$ to an SI voltage cancels the black hole mass when both $Q$ and $r_+$ are expressed through the same geometric mass scale. More importantly, multiplying that voltage by an arbitrarily chosen fragment charge would still give only a kinematic upper bound. Synchrotron and curvature radiation, inverse Compton losses, collisions, pair cascades, screening, self-force, and reaction forces are outside the model. Any astrophysical energy estimate must introduce a concrete plasma environment and solve the associated electrodynamics.}
\endgroup

\subsection{What is genuinely model dependent?}
The comparison in Fig.~\ref{fig:benchmarks} separates three statements. The existence of negative canonical energy for an oppositely charged particle and the leading factor $|q_2|\Delta\Phi_+$ are inherited from the charged-horizon mechanism represented by RN. The nonlinear and $f(R,T)$ coefficients change this scale by shifting $x_+$ and by reshaping $\cF$ outside the horizon. The sign of $\beta$ additionally changes the global endpoint of an outward trajectory. Hence the robust new information is a controlled geometric and causal map of the bound in this particular solution family, rather than a claim of a new microscopic extraction channel.

\section{Conclusions}\label{sec:conclusions}
\rev{We have reformulated charged energy extraction in the purely electric Euler--Heisenberg type $f(R,T)$ solution as a hierarchy of conditional statements. The general local bound retains the infalling fragment's radial velocity and angular momentum; the conventional $w_2=j_2=0$ choice is its upper envelope. A co-moving split supplies a stricter benchmark that conserves four momentum exactly. The horizon structure is classified with the event and cosmological roots distinguished and with extremality fixed by $\cF=\cF'=0$.}

The numerical comparison with RN, Einstein--Euler--Heisenberg type, and $f(R,T)$--Maxwell limits shows that the principal variation of the local envelope is the deformation-induced shift of the branch-referenced horizon potential $\Delta\Phi_+$. Since the electric field itself remains Coulombic, this effect is geometry mediated. \rev{In the dS static patch, $\Delta\Phi_+=q(1/x_+-1/x_c)$; in the flat and AdS branches, $\Delta\Phi_+=q/x_+$. Global propagation is not universal: the flat branch can admit escape to infinity, the dS branch admits propagation only across the static patch to the cosmological horizon, and a finite energy massive fragment in the AdS branch has an outer turning point.}

\rev{All efficiencies reported here are fixed background test particle bounds. They do not include self-force, radiation reaction, pair creation, plasma screening, charge depletion, or electromagnetic backreaction. }

\end{document}